\documentclass[aps,amsmath,twocolumn,superscriptaddress,prl,showpacs]{revtex4}

\usepackage{graphicx}

\begin{document}

\preprint{}

\title{Theory of trapped polaritons in patterned microcavities}

\keywords{Excitons, polaritons, microcavities, quantum confinement, quantum dots.}

%% Please do not enter footnotes or \inst{}-notes into the optional
%% argument of the author command. The optional argument will go into
%% the header.  If there is only one address the marker \inst{x} may be
%% omitted.

%% Information for the first author.
\author{Pierre Lugan}
\affiliation{Institut de Th\'eorie des Ph\'enom\`enes Physiques, Ecole
Polytechnique F\'ed\'erale de Lausanne (EPFL), CH-1015 Lausanne, Switzerland}
\author{Davide Sarchi}
\affiliation{Institut de Th\'eorie des Ph\'enom\`enes Physiques, Ecole
Polytechnique F\'ed\'erale de Lausanne (EPFL), CH-1015 Lausanne, Switzerland}
\author{Vincenzo Savona}
\email[]{vincenzo.savona@epfl.ch}
\affiliation{Institut de Th\'eorie des Ph\'enom\`enes Physiques, Ecole
Polytechnique F\'ed\'erale de Lausanne (EPFL), CH-1015 Lausanne, Switzerland}
%%
%%    \dedicatory{This is a dedicatory.}
\begin{abstract}
We consider the system of a quantum well embedded in a planar semiconductor 
microcavity with a shallow circular mesa patterned on top of the cavity 
spacer. For this system we develop the linear coupling theory of polaritons. 
We then compute polariton eigenstates and the corresponding optical spectrum. 
The theory predicts the existence of laterally confined polariton states with 
a discrete energy spectrum, as well as continuum states above the finite mesa 
potential barrier.
\end{abstract}

\pacs{71.36.+c,71.35.Lk,42.65.-k}

\maketitle

Polaritons in planar semiconductor microcavities (MCs) are quasi-particles 
ideally characterized by free motion in two-dimensions. The most remarkable 
experimental signature of free two-dimensional motion is the continuous 
energy-momentum dispersion curve \cite{HoudrePRL1994,LangbeinJPCM2004}. More 
direct evidence was recently provided by measurements of the polariton motion 
along the plane \cite{LangbeinICPS2002,FreixanetPRB2000}, that showed 
ballistic propagation over hundreds of $\mu$m when polariton wave-packets were 
resonantly excited at large momentum. 

The free particle picture, however, does not hold exactly. The structural 
disorder in the system breaks the translational invariance along the plane and 
produces localized polariton states. The polariton localization length in a 
GaAs-based high-quality sample was estimated in the range of a few tens of 
$\mu$m close to the band bottom, by measuring the momentum broadening of the 
dispersion curve \cite{LangbeinPRL02}, whereas recent measurements of 
spatially-resolved photoluminescence from a ZnSe-based sample 
\cite{RichardPRB2005} directly showed the spatial pattern of localized 
polariton states. Presumably, polariton localization occurs mainly due to 
disorder in the photonic structure. The very small polariton effective mass, 
as compared to the exciton mass, is indeed expected to produce {\em motional 
narrowing} \cite{WhittakerPRL1998} that strongly suppresses the effect of 
exciton disorder on the polariton states. In the MC, on the other hand, small 
thickness fluctuations can produce sizeable local variations of the photon 
resonant energy, that are expected to affect the polariton motion.

The idea of producing a polariton trap by a local variation of the MC 
thickness was recently put forward \cite{EldaifAPL2006} for engineering a 
``polariton quantum box'', namely a photonic structure able to confine 
polaritons in the three spatial directions. Given the very light polariton 
mass, of the order of $10^{-5}$ exciton masses, the trap size required for 
producing a sizeable energy quantization is 300 times larger than in the case 
of bare excitons. Polariton quantum boxes as large as 10 $\mu$m can thus be 
designed. Compared to quantum dots \cite{BimbergBOOK1999}, these photonic 
structures are easier to fabricate, allowing perfect control over size, shape 
and positioning. Addressing a single quantum box with optical spectroscopy 
would also become a trivial experimental task. Another advantage with respect 
to extended polaritons is the reduced spectral linewidth -- as is the case for 
quantum dots \cite{GammonScience1996,BorriPRL2001} -- due to the fact that a 
spatially localized state samples a homogeneous portion of the disordered MC 
plane. Finally, trapped polaritons still retain their quasi-bosonic character, 
contrarily to electron-hole states in quantum dots. Indeed, in order for the 
Pauli exclusion to break the bosonic behaviour, it is required that the {\em 
exciton} density approach its saturation value \cite{SchmittrinkPRB1985}. 
Populating a single confined state with a few polaritons still keeps the 
exciton density well below this limit. This quite unique situation, thanks to 
the discrete energy spectrum, would allow Bose-Einstein condensation of 
polaritons, in contrast to the ideally homogeneous two-dimensional case, for 
which the Hohenberg-Mermin-Wagner theorem \cite{HohenbergPR1967} strictly 
forbids the occurrence of this quantum phase transition.

\begin{figure}[h!]
\includegraphics[width=0.40 \textwidth]{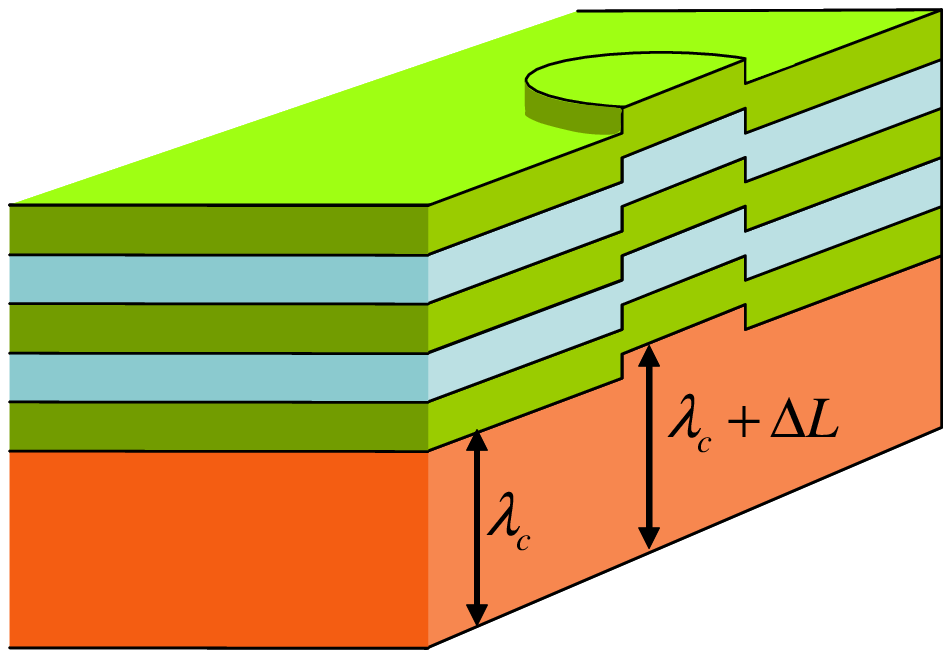}
\hfil
\includegraphics[width=0.50 \textwidth]{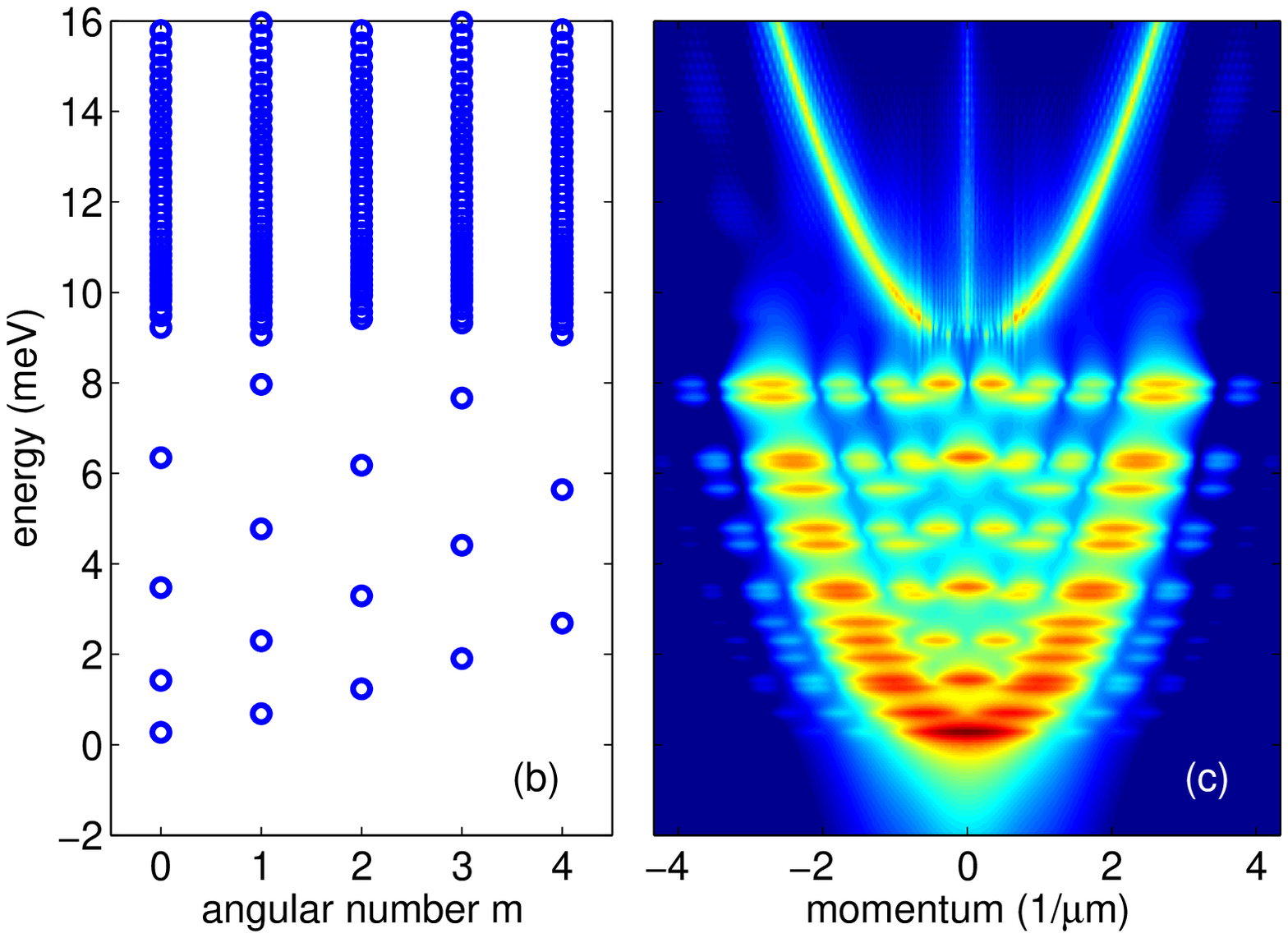}
\caption{(a) Sketched cross section of the mesa structure. The various lengths and the number of layers are not in scale. (b) Computed energy eigenvalues for the photon modes of a circular mesa ($D=8.6$ $\mu$m, $\hbar\Delta\omega_c=-9$ meV). (c) Intensity plot of the photon spectral density (log scale, 4 decades from blue to red).}
\label{fig1}
\end{figure}

Let us assume a $\lambda_c$-thick MC with distributed Bragg mirrors (DBRs). 
Following the scheme proposed by El Da\"if {\em et al.} \cite{EldaifAPL2006}, 
in order to create a polariton trap, the thickness of the MC spacer is made 
slightly larger by an amount $\Delta L$ within a limited region of the MC 
plane that we call {\em mesa}, as sketched in Fig. \ref{fig1} (a). This local 
variation is obtained by etching the mesa pattern on top of the cavity spacer 
and then growing the top DBR. To an increased cavity thickness corresponds a 
local decrease of the resonant photon-mode energy. As the photon couples to 
the exciton, forming a polariton mode, the mesa acts as a polariton trap. We 
point out that only the MC spacer thickness is varied, while the optical 
thickness of the DBR layers is assumed everywhere equal to $\lambda_c/4$. 
Therefore, within the mesa the DBRs are slightly detuned with respect to the 
spacer thickness. This makes fabrication easier and implies a less pronounced 
dependence of the local mode energy on $\Delta L$. In fact, the lowest MC 
resonant fequency is obtained as the solution of $(n_c/c)[(\omega-
\omega_c)L_c+(\omega-\omega_m)L_{DBR}]=0$, where $L_c$ is the local cavity 
thickness, $\omega_c$ and $\omega_m$ are respectively the resonant frequency 
of the cavity spacer and of the DBRs, and 
$L_{DBR}=(\lambda_c/2)n_1n_2/[n_1(n_1-n_2)]$ is the field penetration length 
in the DBRs, expressed in terms of the two DBR refraction indices $n_1>n_2$. 
By replacing $L_c=\lambda_c+\Delta L$, $\omega_m=(c/n_c)2\pi/\lambda_c$, and 
$\omega_c=(c/n_c)2\pi/(\lambda_c+\Delta L)$, we obtain $\Delta\omega_c=-
\omega_m\Delta L/(\lambda_c+L_{DBR})$. For ideal DBRs tuned to the spacer 
frequency, instead, $\Delta\omega_c=-\omega_m\Delta L/\lambda_c$. This weaker 
dependence on $\Delta L$ has thus the further advantage of allowing finer 
control over the energy offset of shallow mesas, for which a $\Delta L$ of 
only a few nm must be obtained in the fabrication process. For typical 
GaAs/AlAs DBRs, $L_{DBR}\simeq3\lambda_c$. Given $\Delta L=6$ nm in Ref. 
\cite{EldaifAPL2006}, the previous expression predicts a mesa energy offset of 
$\hbar\Delta\omega_c=-9$ meV, in agreement with the optical characterization 
of the sample.

For a shallow mesa of lateral extension larger than the optical wavelength, we 
can safely assume that the electromagnetic modes at in-plane position 
$\pmb\rho$ are locally equivalent to those of a planar microcavity:
\begin{equation}
{\bf E}({\bf r})={\bf E}({\pmb \rho})\exp(ik_z({\pmb\rho})z)\,,
\label{ansatz}
\end{equation}
where $k_z({\pmb\rho})$ is piecewise constant. 
Neglecting border effects at the mesa contour, Maxwell equations give
\begin{equation}
\nabla^2_{\rho}{\bf E}({\pmb \rho})+\left(\frac{\omega^2}{c^2}\epsilon_0-k_z^2({\pmb\rho})\right){\bf E}({\pmb\rho})=0\,,
\label{maxwell2d}
\end{equation}
where $\epsilon_0$ is the spacer dielectric constant. Outside the mesa, the MC 
resonance is $k_z=2\pi/\lambda_c$. Inside the mesa we can relate the offset 
$\Delta k_z$ to the energy offset $\Delta \omega_c$, as $\Delta 
k_z=\sqrt{\epsilon_0}\Delta \omega_c/c$. Eq. (\ref{maxwell2d}) is then solved 
in cylindrical coordinates, assuming a circular mesa of diameter $D$. The 
eigenmodes are therefore expressed as ${\bf 
E}({\pmb\rho})=U_{nm}(\rho)\exp(im\phi)$, where $n=0,1,\ldots$ and $m=-
n,\ldots,n$ are the radial and angular mode numbers respectively. The mode 
eigenenergies and the corresponding energy-momentum spectral function are 
plotted in Fig. \ref{fig1} (b) and (c). A discrete energy spectrum appears, 
whose modes show a flat extended signature in momentum space, reflecting their 
spatial confinement. At energy above the 9 meV barrier, we find a continuum of 
states whose energy-momentum signature practically coincides to that of a 
planar MC. This result already suggests that the structure is able to confine 
photons in the three spatial directions.

We similarly express the exciton center-of-mass wave function in terms of 
Bessel functions of the first kind, 
$\psi_{nm}({\pmb\rho})=N_{nm}J_m(\kappa_{nm}\rho)\exp(im\phi)$, where 
$\kappa_{nm}$ are the exciton eigen-momenta and $N_{nm}$ is a normalization 
constant. These are the modes of a free particle, as the exciton motion is not 
affected by the mesa structure. By introducing Bose operators $\hat{A}_{nm}$ 
and $\hat{B}_{nm}$ for photon and exciton modes respectively, the linear 
exciton-photon Hamiltonian can be finally expressed in second quantization as
\begin{eqnarray}
H&=&\sum_m\left[\sum_n\hbar\omega^{(ph)}_{nm}\hat{A}^\dagger_{nm}\hat{A}_{nm}+\sum_n\hbar\omega^{(exc)}_{nm}\hat{B}^\dagger_{nm}\hat{B}_{nm}\right.\nonumber\\
&+&\left.\left(\sum_{nn^\prime}\hbar\Omega^{(m)}_{nn^\prime}\hat{A}^\dagger_{nm}\hat{B}_{n^\prime m}+ \mbox{h.c.}\right)\right]\,,
\label{hamilt}
\end{eqnarray}
where $\omega^{(ph)}_{mn}$ and $\omega^{(exc)}_{mn}$ are the eigenenergies of the photon and of the (free) exciton modes. As required by symmetry, the angular number $m$ is conserved in the coupling. The energies $\hbar\Omega^{(m)}_{nn^\prime}$ are expressed in terms of the Rabi splitting of the planar cavity $\hbar\Omega_R$ and of exciton-photon overlap integrals
\begin{equation}
\Omega^{(m)}_{nn^\prime}=2\pi\Omega_R\int d\rho\,\rho\, U^*_{nm}(\rho)N_{n^\prime m}J_m(\kappa_{n^\prime m}\rho)\,.
\label{overlap}
\end{equation}

\begin{figure}[h!]
\includegraphics[width=0.50 \textwidth]{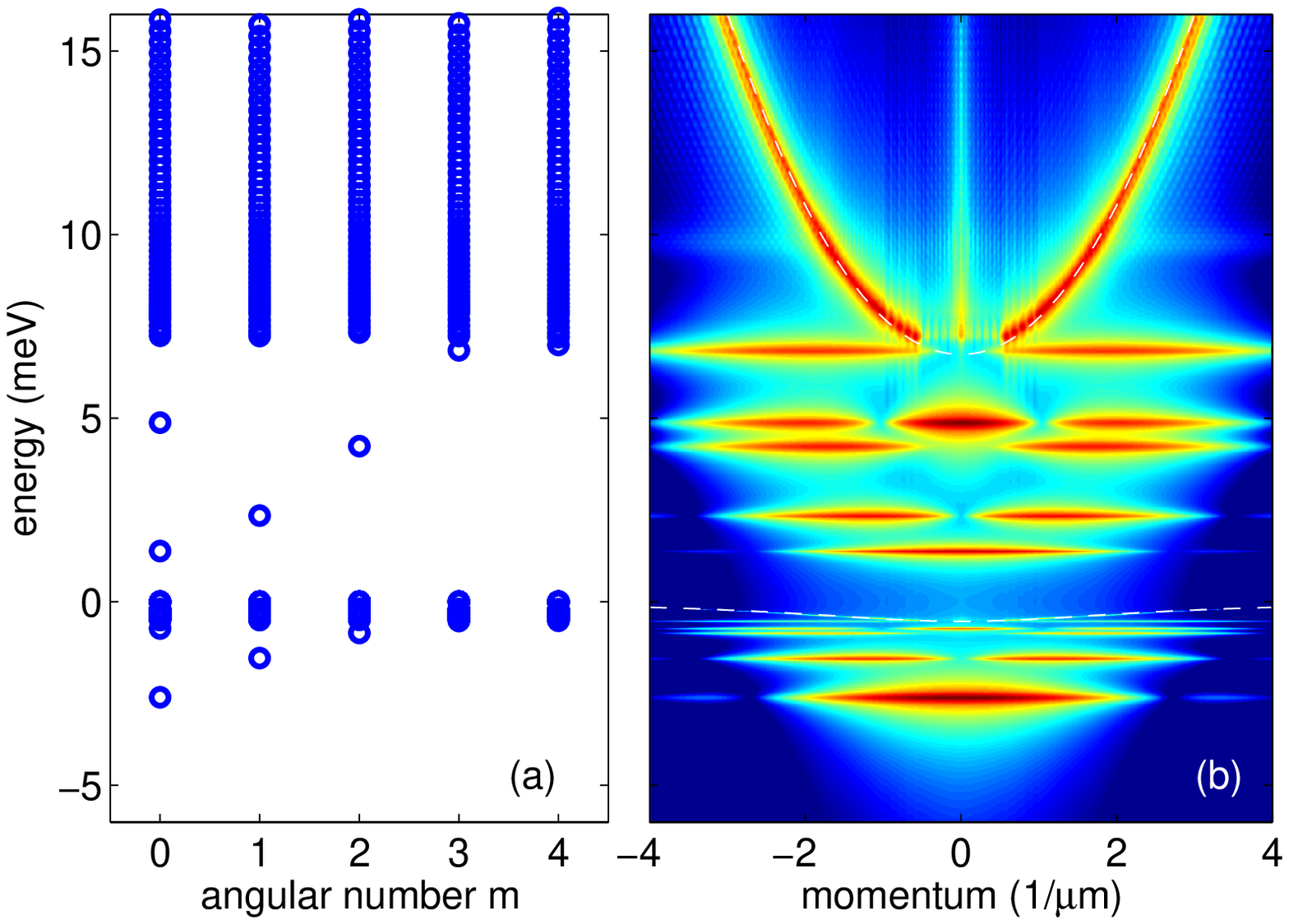}
\includegraphics[width=0.50 \textwidth]{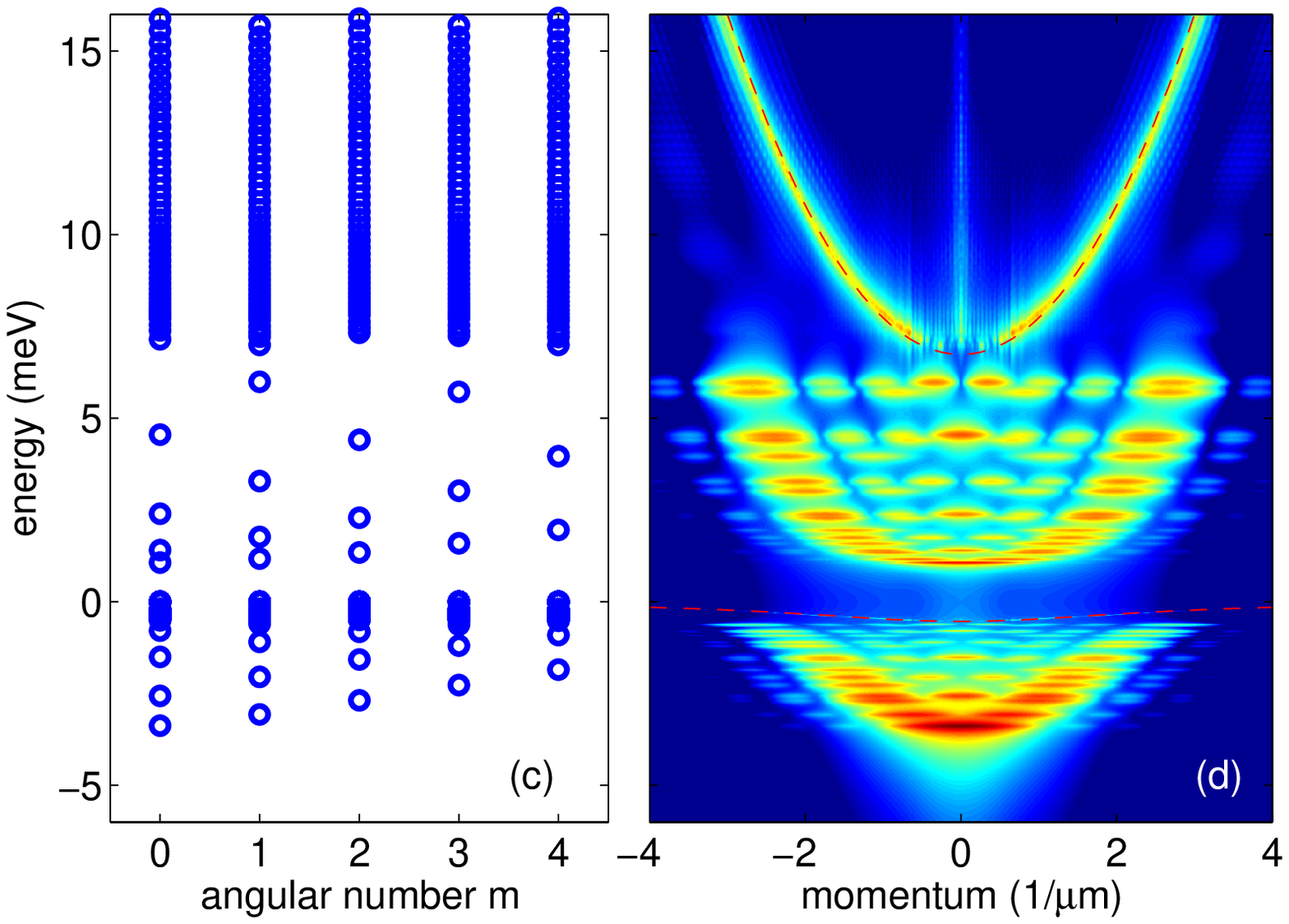}
\caption{(a) Computed polariton eigenvalues for $D=3.5$ $\mu$m. (b) Intensity plot of the polariton energy-momentum spectral function for $D=3.5~\mu$m (log scale, 4 decades from blue to red). (c) and (d): same as (a) and (b) for $D=8.6$ $\mu$m. The dashed lines indicate the bare polariton dispersion of the planar MC corresponding to the mesa barrier.}
\label{fig2}
\end{figure}

For a planar geometry, momentum conservation implies a one-to-one coupling 
between exciton and photon modes. Here, on the contrary, no selection rule on 
the radial quantum number $n$ exists. For the numerical solution we therefore 
choose to retain only a finite number of cavity modes $N_c$ and exciton modes 
$N_x$ for each value of $m$. The resulting polariton modes are obtained by 
numerical diagonalization of the $(N_c+N_x)\times(N_c+N_x)$ matrix obtained 
from the Hamiltonian (\ref{hamilt}). Polariton eigenvalues obtained for 
$D=3.5$ $\mu$m and $D=8.6$ $\mu$m are plotted in Fig. \ref{fig2} (a) and (c) 
respectively. A detuning of $+6.5$ meV for the planar cavity mode with respect 
to the exciton was assumed, in order to bring the lowest confined photon modes 
in resonance with the bare exciton. The result in Fig. \ref{fig2} (a) and (c) 
brings clear evidence of a discrete energy spectrum, followed by a continuous 
spectrum at higher energy, both for the lower and for the upper polariton. For 
the smaller mesa, only three lower polariton and four upper polariton states 
are present in the discrete spectrum, with energy spacings in the order of a 
few meV. The spectra recently measured by El Da\"if {\em et al.} on mesas of 
comparable size show spectral features in agreement with the present model. 
The polariton operators obtained from the diagonalization of (\ref{hamilt}) 
are expressed as 
$\hat{P}_{nm}=\sum_{n^\prime}(X_{nm}^{n^\prime}\hat{B}_{n^\prime 
m}+W_{nm}^{n^\prime}\hat{A}_{n^\prime m})$. Each polariton mode exhibits an 
angular emission pattern according to its photon component in momentum space, 
defined as $I_{nm}({\bf k})=|\langle{\bf k}|\hat{P}_{nm}^\dagger|0\rangle|^2$, 
that is easily computed from the model. By assuming for each mode a lorentzian 
energy spectrum, we can finally compute energy-momentum spectral function, as 
shown in Fig. \ref{fig2}(b) and (d). In the discrete part of the spectrum, 
polariton modes present a flat, broad energy-momentum signature, which 
corresponds to the Fourier transform of spatially confined states (not shown). 
The continuous part of the spectrum, on the other hand, simply corresponds to 
the dispersion of free two-dimensional polaritons. These are the scattering 
states above the finite energy barriers of the potential formed by the mesa. 
Correspondingly, the energy-momentum dispersion is well defined, with a 
negligible broadening in k-space. We point out that the discrete modes for 
$D=8.6~\mu$m are less extended in momentum space than those for $D=3.5~\mu$m, 
forming a quasi-continuum that mimicks the energy-momentum dispersion pattern 
of 2-D polaritons. For diameters larger than 20 $\mu$m, the simulations show a 
spectrum practically identical to that of 2-D polaritons in a MC of thickness 
$\lambda_c+\Delta L$.

In conclusion, we have developed a theoretical model of polaritons in 
patterned mesa structures. The model predicts a discrete energy spectrum of 
confined polariton states both for the upper and the lower branch, each 
coexisting with a continuous spectrum of extended states above the trap 
barrier. The confined states are extended in momentum space -- a feature that 
should be detected in angle-resolved spectroscopy. Polariton quantum boxes are 
interesting in several respects. The quantum confinement is produced in a way 
that is not as invasive as in micropillars \cite{DasbachPRB2001}, which seem 
to considerably reduce the quality factor of the photonic structure, 
presumably by opening lateral emission channels. Confined polaritons represent 
a quite unique system of quantum-confined interacting bosons. They can undergo 
parametric scattering processes -- due to the mutual Coulomb interaction --
that have recently been exploited for producing nonclassical states of 
polaritons \cite{SavastaPRL2005}. In quantum boxes, momentum conservation no 
longer holds and parametric processes between quantum confined states become 
allowed. This might be exploited for designing a quantum optical device. 
Recently, rigurous proof was given, that a discrete energy spectrum allows 
Bose-Einstein condensation \cite{LauwersJPA2003}, otherwise impossible in 
two-dimensions \cite{HohenbergPR1967}. Polariton quantum boxes, could therefore be 
the ideal system for observing the long sought Bose-Einstein condensation of 
polaritons.

We are grateful to B. Deveaud, W. Langbein and R. Zimmermann for the fruitful 
collaborations. Financial support by the Swiss National Science Foundation 
(project No. 620-066060) is also acknowledged.

\end{document}